\documentstyle[osa,aps,prb,epsfig]{revtex}
\begin{document}

\twocolumn[\hsize\textwidth\columnwidth\hsize\csname
      @twocolumnfalse\endcsname

\title{Fabrication and Electrical Properties of Pure VO$\bf_{2}$ Phase Films}

\author{B. G.~{\textsc{Chae}}, D. H.~{\textsc{Youn}}, H. T.~{\textsc{Kim}},
S.~{\textsc{Maeng}},\\ and K. Y.~{\textsc{Kang}}}

\address{Basic Research Laboratory, ETRI, Daejeon 305-350,
Republic of Korea}

\maketitle{}

\begin{abstract}
We have grown VO$_{2}$ thin films by laser ablation for electronic
device applications. In obtaining the thin films of the pure
VO$_{2}$ phase, oxygen partial pressure is a critical parameter
because vanadium oxides have several phases with differing oxygen
concentration. It is found that pure VO$_{2}$ films are
epitaxially grown on $\alpha$-Al$_{2}$O$_{3}$ substrate in the
narrow ranges of 55-60 mTorr in an Ar + 10\% O$_{2}$ ambient, and
that mixed phase films are synthesized when the deposition
pressure slightly deviates from the optimum pressure. The (100)
oriented VO$_{2}$ films undergo an abrupt metal-insulator
transition (MIT) with resistance change of an order of 10$^{4}$ at
338 K. In the films of mixed phases, the small change of the
resistance is observed at the same temperature. The
polycrystalline films grown on SiO$_{2}$/Si substrate undergo
broaden change of the resistance. Furthermore, the abrupt MIT and
collective current motion appearing in metal are observed when the
electric field is applied to the film.
\\
\\
\end{abstract}
]

\section{Introduction}

 Oxides of vanadium undergo a metal-insulator transition (MIT)
at a critical temperature.\cite{1,2,3} Among these oxides,
vanadium dioxide VO$_2$ has been attracted much attention because
its transition temperature is near room temperature, 340
K.\cite{4,5} The electrical-resistivity change of an order of
10$^{4}$-$10^{5}$ and the abrupt infrared-transmission change are
observed at the transition temperature, which makes it useful for
application to electrical and optical switching devices.\cite{6,7}

 The properties of the phase transition of VO$_{2}$ films strongly
depends on the nature of the crystal structure and their
stoichiometry, including the film orientation, oxygen content, and
residual interface strain. In particular, the oxygen partial
pressure is very sensitive in growing the films with a pure
VO$_{2}$ phase. Thus, it is important and necessary, although not
easy, to obtain pure phase films for better properties of
metal-insulator transition.

  To date, there have been various methods used for the
successful deposition of VO$_{2}$ thin films, such as chemical
vapor deposition,\cite{8} reactive sputtering,\cite{9} and laser
ablation.\cite{5,6} Pulsed laser deposition was proven to be an
excellent method, especially in growing oxide thin films.

  In this paper, we deposit
(100) oriented VO$_{2}$ thin films on sapphire substrate by laser
ablation, with varying the oxygen partial pressure. The
polycrystalline films are grown on the SiO$_{2}$/Si substrate;
detailed properties will be presented in another paper. Phases of
the vanadium oxide films are extensively investigated by means of
observations of crystal structure and of the resistance change
with temperature. Furthermore, metal-insulator transition induced
by the electric field is observed.

\section{Experiments}

 The VO$_{2}$ thin films were successfully grown on $\alpha$-Al$_{2}$O$_{3}$
($\bar{1}$012) and SiO$_{2}$/Si substrates by laser ablation with
a vanadium metal target in an ambient of a partially filled oxygen
and argon. The KrF excimer laser (Lambda-Physik, Compex 205) with
a wavelength of 248 nm was used to ablate the rotating metal
target. An energy density of 1$\sim$2 J/cm$^{2}$ was focused on
the target surface at a repetition rate of 5 Hz. The distance
between the target and the substrate was 5 cm. The detailed
conditions for the preparation of the vanadium oxide films are
described in Table 1. Prior to deposition, the substrates were
cleaned with a formal process to remove residual contaminants on
their surfaces. The chamber was evacuated down to the base
pressure as low as 10$^{-6}$ Torr. Argon and oxygen gases were
filled by adjusting the gas-flow meter. The substrate temperature
was kept at 450 $\rm^{\circ}$C during the deposition process. The
film growth was carried out at a pressure between 50 and 200
mTorr. The target was pre-ablated with a shutter for the several
minutes to clean the target surface. The partial pressure of
oxygen was the most critical variable in obtaining the pure
VO$_{2}$ phase, which was controlled by the working pressure
containing 10\% oxygen in an argon atmosphere. The deposition rate
of VO$_{2}$ films was estimated to be about 0.39 \AA/sec. After
deposition, the substrates were slowly cooled to the room
temperature under the same deposition atmosphere as that used for
growth.

 The crystalline structure of deposited films was analyzed
by X-ray diffraction (XRD)and high energy electron diffraction
(RHEED). The surface morphology and the film composition were
observed by scanning electron microscopy (SEM) and secondary ion
mass spectrometry (SIMS), respectively. The resistivity of the
films was measured using four-probe method. To observe the
metal-insulator transition with respect to the electric field, the
Au/Cr electrodes were patterned on VO$_{2}$ thin films by the
lift-off method.

\section{Results and Discussion}

 Figure 1 shows X-ray diffraction (XRD) patterns of vanadium oxide
films deposited on $\alpha$-Al$_{2}$O$_{3}$ ($\bar{1}$012)
substrates in different ambient pressures. The film orientation
strongly depends upon partial oxygen contents in the deposition
ambient. Although vanadium oxides have several phases, no peaks of
phases exist other than those of three phases VO$_{2}$,
V$_{6}$O$_{13}$, and V$_{2}$O$_{5}$. Figure 1(a) shows that the
film synthesized in a pressure of more than about 120 mTorr, is
highly (00l) oriented to V$_{2}$O$_{5}$. The film deposited in 70
mTorr has mixed phases of VO$_{2}$, V$_{6}$O$_{13}$, and
V$_{2}$O$_{5}$, as shown in Fig. 1(b). In particular, the (00l)
peaks of V$_{6}$O$_{13}$ phase were distinct. Figure 1 (c) shows
only one peak corresponding to the (200) peak of the monoclinic
phase, which indicates the highly preferred orientation of the
film. This is in agreement with the results reported by other
groups.\cite{5,6} Thus, VO$_{2}$ films are grown only in ambient
pressures in the range of 55-60 mTorr. Figure 2 shows an XRD
pattern of VO$_{2}$ thin film grown on SiO$_{2}$/Si substrates.
The film was deposited in the same growth conditions on sapphire
except for at the deposition temperature of 460 $\rm^{\circ}$C.
The VO$_{2}$ phase peaks are formed, which means that the VO$_{2}$
film could be grown even on amorphous SiO$_{2}$.

 Figure 3 shows the SEM image and the RHEED pattern of the VO$_{2}$ thin
films grown on sapphire substrates. As shown in Fig 3(a), the film
surface are formed smoothly with no particulate and grains are
densely packed. The average grain size of the film is about 60 nm.
A typical RHEED pattern for the VO$_{2}$ film is appeared in Fig
3(b). The pattern was obtained with an electron energy of 15 keV
at a room temperature. The arranged spots are observed although
the streaks and Kikuchi lines related to uniform crystal surface
are not appeared. The spotted pattern is formed when the films
have single crystal structure. Therefore the VO$_{2}$ thin films
on sapphire were grown epitaxial. Figure 4 shows SIMS results of
VO$_{2}$ samples. The element distribution along a depth of the
film from a surface is plotted. Vanadium ion for both films grown
on sapphire and SiO$_{2}$/Si substrates deeply penetrates into the
substrate. The interface region is largely formed. Silicon ion
also diffuses into the VO$_{2}$ film in Fig. 5(b). This large
interdiffusion of elements at the interface may be an obstacle to
the growth of pure phase films.

 Figure 5(a) shows the temperature dependence of the
electrical resistance for vanadium oxide films grown on sapphire
in the range of 50-70 mTorr as a function of temperature. The
resistance measurement was carried out in a cryostat where the
samples are cooled and heated by liquid nitrogen and a heating
source. The films grown at 55 and 60 mTorr show abrupt resistance
change of in the order of 3$\times$10$^{4}$ near a critical
temperature, $T_{c}$$\approx$338 K, which is a structural
MIT\cite{1,2}. This abrupt change with a transition width of about
3 K is comparable to that measured in VO$_{2}$ single crystal. The
electrical resistance below $T_{c}$ exponentially increases with a
decreasing temperature of the film, which is a typical
characteristic of semiconductor. The resistance change near
$T_{c}$ for the film grown at the ambient pressure of 65 mTorr is
smaller than others.

 Figure 5(b) shows that, in a film deposited at 70 mTorr, the resistance
change both below $T_{c}$ and near $T_{c}$ is much smaller. This
property may be attributed to the existence of the mixed phases in
the films. Moreover, a film grown at 50 mTorr shows no distinct
transition although the film has VO$_{2}$ phase from x-ray
analysis. This may be considered due to the oxygen deficiencies
generated during the deposition with a low oxygen partial
pressure. As a result of our experiments, it is found that the
pure VO$_{2}$ thin films are obtained at the narrow process window
within the range of 5-10 mTorr. It should be noted that the
control of the oxygen content during the deposition is most
important.

 Figure 6 (a) shows the hysteresis of the resistivity
measured by increasing and decreasing the temperature. This
thermal hysteresis is evidence of the abrupt first-order phase
transition. The width of hysteresis is estimated to be about 4 K.
Figure 6 (b) shows the temperature dependence of the electrical
resistance for the VO$_{2}$ film deposited on SiO$_{2}$/Si. It
undergoes a metal-insulator transition near 341 K, which indicates
formation of the VO$_{2}$ phase. There is no abrupt change of
resistivity of the VO$_{2}$ films compared with the
VO$_{2}$/Al$_{2}$O$_{3}$ films, but the resistance exponentially
decreases with an increasing temperature from 340 K and has a
transition width of 10 K. These characteristics are resulted from
that the film is polycrystalline, as shown in the XRD pattern of
Fig. 2.

 Figure 7 shows the MIT induced by the electric field in a
VO$_{2}$ film grown on sapphire substrate. The schematic diagram
of a two terminal structure is displayed in the inset figure. The
Au/Cr electrodes to apply the electric field was prepared using
the rf-sputtering technique, and patterned into a length of 5
$\mu$m and a width of 25 $\mu$m by the lift-off method. The
experiment was conducted at room temperature and the maximum
current in the instrumentation was limited to prevent the film
from damage suffered from a large current flow in advance; the
current compliance was 5 mA. The current density increases with an
increasing applied voltage below point $A$ in Fig. 7. At the
indicated point $A$, an abrupt jump is shown; this has been
reproducibly observed, and is not interpreted as dielectric
breakdown.\cite{10,11} At point $B$, the current density is about
3 $\times$ 10$^{5}$ A/cm$^{2}$; this order is given due to
compliance, if there is no compliance, the current density is
larger than this value. This is regarded as collective current
motion appearing in metal. Thus, the jump is the abrupt MIT. The
mechanism of the jump is given in a previous paper\cite{12}
because it is outside the scope of this paper.

\section{Conclusions}

We successfully deposited pure VO$_{2}$ thin films on the sapphire
by laser ablation. The working pressure to obtain the pure
VO$_{2}$ phase is near 60 mTorr in an ambient of Ar + 10\%
O$_{2}$. The VO$_{2}$ films on sapphire underwent an abrupt MIT at
$T_{c}$=338 K with resistance change of an order of 10$^{4}$. The
MIT and the collective current motion are observed when the
electric field is applied to the film.

\begin{figure}
\vspace{5.0cm}
\centerline{\epsfysize=9cm\epsfxsize=6.5cm\epsfbox{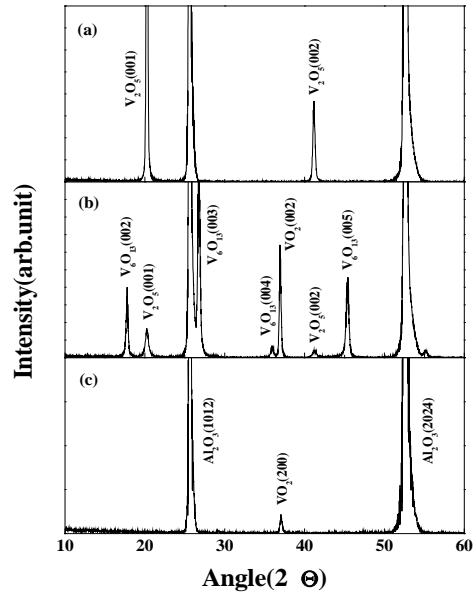}}
\vspace{0.0cm} \caption{X-ray diffraction patterns for the
VO$_{2}$ thin films grown on $\alpha$-Al$_{2}$O$_{3}$
($\bar{1}$012) in various pressures.} \label{f1}
\end{figure}

\begin{figure}
\vspace{-1.0cm}
\centerline{\epsfysize=9cm\epsfxsize=6.5cm\epsfbox{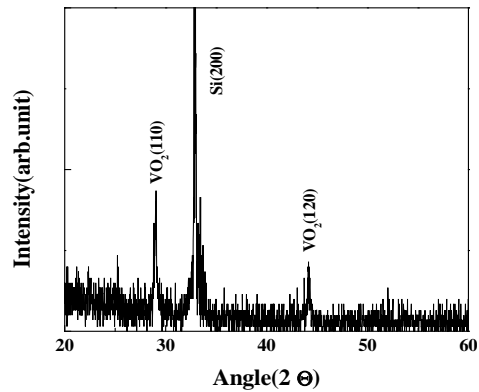}}
\vspace{-1.0cm} \caption{X-ray diffraction patterns for the
VO$_{2}$ thin films grown on SiO$_{2}$/Si substrates at 460
$\rm^{\circ}$C.} \label{f2}
\end{figure}

\begin{figure}
\vspace{-4.0cm}
\centerline{\epsfysize=11cm\epsfxsize=8.5cm\epsfbox{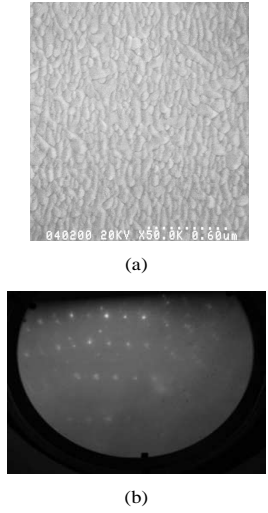}}
\vspace{-1.5cm} \caption{(a) SEM images and (b) RHEED pattern of
VO$_{2}$ thin films grown on Al$_{2}$O$_{3}$ substrate. The films
are deposited at 60 mTorr.} \label{f3}
\end{figure}

\begin{figure}
\vspace{0.0cm}
\centerline{\epsfysize=11cm\epsfxsize=8.5cm\epsfbox{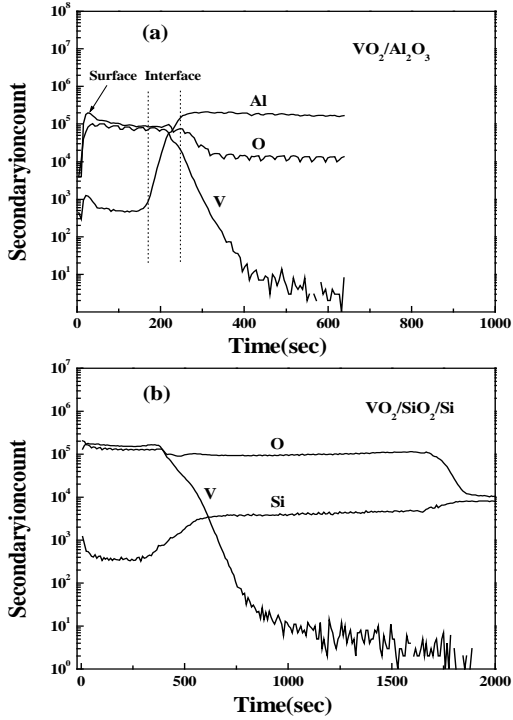}}
\vspace{0.0cm} \caption{SIMS data of VO$_{2}$ thin film grown on
(a) sapphire and (b) SiO$_{2}$/Si substrates.} \label{f4}
\end{figure}

\begin{figure}
\vspace{-1.0cm}
\centerline{\epsfysize=11cm\epsfxsize=8.5cm\epsfbox{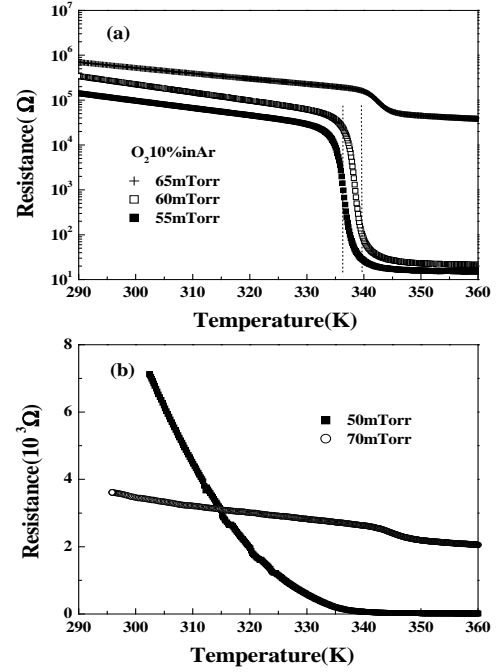}}
\vspace{-0.5cm} \caption{Changes of electrical resistance as a
function of temperature for vanadium oxide films grown on
$\alpha$-Al$_{2}$O$_{3}$ substrate.} \label{f5}
\end{figure}

\begin{figure}
\vspace{-1.0cm}
\centerline{\epsfysize=11cm\epsfxsize=8.5cm\epsfbox{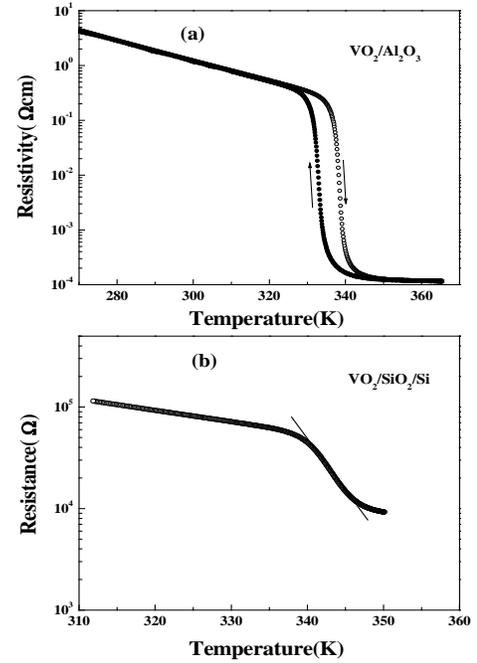}}
\vspace{-0.5cm} \caption{(a) Thermal hysteresis of resistivity
with temperature for VO$_{2}$ grown on sapphire substrate. (b)
Changes of electrical resistance as a function of temperature for
VO$_{2}$ grown SiO$_{2}$/Si substrate.} \label{f6}
\end{figure}

\begin{figure}
\vspace{0.0cm}
\centerline{\epsfysize=11cm\epsfxsize=8.5cm\epsfbox{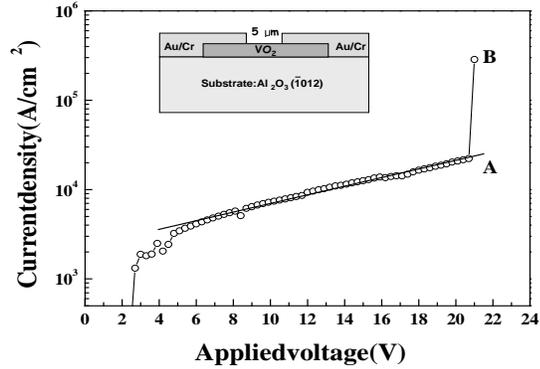}}
\vspace{-3.5cm} \caption{Changes of currents by application of the
electric field to VO$_{2}$ thin films. Inset: Schematic diagram
for observance of effects of electric field on metal-insulator
transition of VO$_{2}$ thin films.} \label{f7}
\end{figure}

\begin{table}[tb]
\caption{Deposition conditions for preparation of VO$_{2}$ thin
films by laser ablation.}
\label{t1}

\begin{center}

\begin{tabular}{cc}
Atomsphere & Ar + O210\% \\
Laser power & 200 mJ \\
Repetition rate & 5 Hz \\
Target & V metal target \\
Substrate & $\alpha$-Al$_{2}$O$_{3}$ ($\bar{1}$012)and
SiO$_{2}$/Si \\
Initial pressure & below 10$^{-6}$ Torr \\
Deposition pressure & 50$\sim$200 mTorr \\
Substrate-target distance & 5 cm \\
\end{tabular}

\end{center}

\end{table}

\end{document}